\documentclass[prl,showpacs,aps,twocolumn]{revtex4}
\usepackage{graphicx}
\newcommand{\beqn}{\begin{equation}}
\newcommand{\eeqn}{\end{equation}}
\newcommand{\bet}{\begin{tabular}}
\newcommand{\ent}{\end{tabular}}

\begin{document}
\title{Identification and characterization of icosahedral metallic nanowires }

\author{Samuel Pel\'aez$^1$, Carlo Guerrero$^2$, Ricardo Paredes$^{3}$, Pedro A. Serena$^1$ and Pedro \ Garc\'{\i}a-Mochales $^4$}

\address{$^1$ Instituto de Ciencia de Materiales de Madrid, Consejo Superior de Investigaciones Cient\'{\i}ficas,
c/ Sor Juana In\'es de la Cruz 3, Campus de Cantoblanco, E-28049-Madrid, Spain}
\address{$^2$ Departamento  de F\'{\i}sica, Facultad Experimental de Ciencias, La Universidad del Zulia, Maracaibo, Venezuela}
\address{$^3$ Centro de F\'{\i}sica, Instituto Venezolano de Investigaciones Cient\'{\i}ficas, Apdo 20632, 
Caracas 1020A, Venezuela}
\address{$^4$ Departamento de F\'{\i}sica de la Materia Condensada, Facultad de Ciencias, 
Universidad Aut\'onoma de Madrid, c/ Francisco Tom\'as y Valiente 7, Campus de Cantoblanco, E-28049-Madrid, Spain}

\date{\today}

\begin{abstract}
We present and discuss an algorithm to identify and characterize the long icosahedral structures (staggered pentagonal nanowires with 1-5-1-5 atomic structure) that appear in Molecular Dynamics simulations of metallic nanowires of different species subjected to stretching. The use of the algorithm allows the identification of pentagonal rings forming the icosahedral structure as well as the determination of its number $n_p$, and the maximum length of the pentagonal nanowire $L_p^m$. The algorithm is tested with some ideal structures to show its ability to discriminate between pentagonal rings and other ring structures. We applied the algorithm to Ni nanowires with temperatures ranging between 4K and 865K, stretched along the [100] direction. We studied statistically the formation of pentagonal nanowires obtaining the distributions of length $L_p^m$ and number of rings $n_p$ as function of the temperature. The $L_p^m$ distribution presents a peaked shape, with peaks locate at fixes distances whose separa-tion corresponds to the distance between two consecutive pentagonal rings. 
\end{abstract}

\pacs{ 02.70.-c, 02.70.Ns, 61.46.Km, 62.23.Hj, 62.25.-g }

\maketitle


\section{Introduction}
\begin{figure}[tb]
\includegraphics[width=8.5 cm]{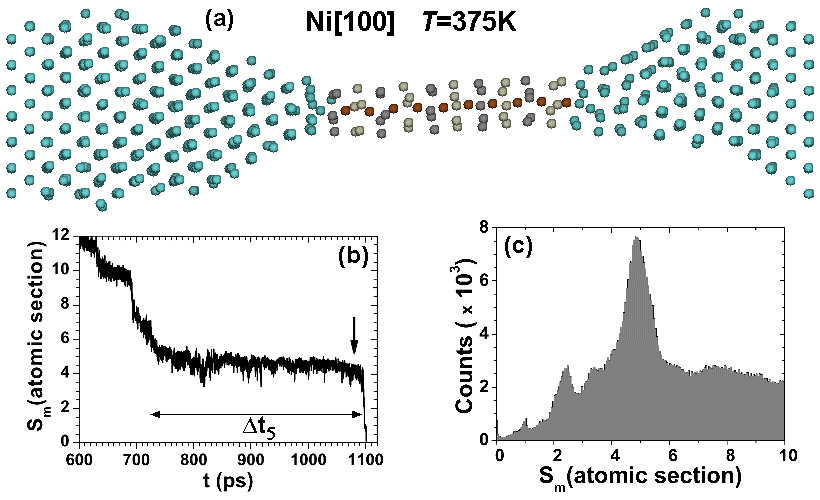}
\caption {\label{fig1} Results from MD simulations of Ni [100] nanowires containing 1029 atoms and subjected to longitudinal stretching at T=375K. (a) Snapshot of a nanowire formed under stretching presenting an icosahedral structure with $n_p=10$ pentagonal rings. (b) Minimum cross section $S_m(t)$ curve corresponding to the simulation depicted in (a) (the arrow points out the time the snapshot was taken). (c) Minimum cross section histogram $H(S_m)$ obtained from tha accumulation of 300 independent breaking events.}
\end{figure}

Icosahedral or pentagonal nanowires are formed by subsequent staggered parallel pentagonal rings (with a relative rotation of $\pi/5$) connected with single atoms, showing a characteristic -5-1-5-1- ordering (see an example in Fig. \ref{fig1}a. Metallic nanowires are of great tech-nological importance due to their properties and potential applications \cite{SerenaBook97,AgraitRev03}. Contrary to monoatomic chains, penta-gonal nanowires are rather robust structures at relatively high temperatures and, therefore, they may consider as a promising candidate for being used as nanodevice components.

Different computational works during the last decade have showed the formation of staggered pentagonal configurations on breaking nanowires of different species \cite{BarnettNature97,MehrezPRB97,GulserenPRL98,GonzalezPRL04,SutrakarJP08,ParkSM06,SenPRB02,GarciaMochalesN08,GarciaMochalesJNM08,GarciaMochalesPSSA08}. The atomic sequence -1-5-1-5- presents a fivefold symmetry with respect the nanowire axis. This symmetry does not correspond to any crystallographic FCC nor BCC structures. The -1-5-1-5- staggered nanowire configuration may be understood in terms of a sequence of interpenetrated icosahedra. This icosahedral symmetry is very common in very small systems due to the large stability and high coordination characterizing such geometry \cite{BulienkovRCBIE01}. 

The formation of staggered pentagonal configurations during the stretching process has been already reported for Na \cite{BarnettNature97} using first principles methods, and for Cu \cite{MehrezPRB97,GonzalezPRL04,SutrakarJP08} and Au \cite{ParkSM06} nanowires with different Molecular Dynamic (MD) approaches. In particular the high stability of the Cu nanowire was confirmed with {\it ab-initio} calculations \cite{SenPRB02}. Pentagonal motives also appear in infinite Al and Pb nanowires obtained from MD simulated annealing methods \cite{GulserenPRL98}. More recently such structures have been reported for stretched Ni nanowires with different crystallographic orientations \cite{GarciaMochalesN08,GarciaMochalesJNM08,GarciaMochalesPSSA08}, and confirmed their stability by {\it ab-initio} simulations \cite{GarciaMochalesJNM08,GarciaMochalesPSSA08}. These pentagonal structures are very stable, with lengths larger than 20 \AA\  and presenting a high plastic deformation under strain. 

In general, the reported pentagonal nanowires have been found for single stretching evens. However, it is well known that the analysis of  nanoscale processes requires the use of statistical approaches since there exists a broad bunch of breaking paths in the nanowire configuration space. Such statistical studies have been only addressed for Ni up to date \cite{GarciaMochalesN08,GarciaMochalesJNM08,GarciaMochalesPSSA08}. It has been shown that [100] and [110] stretching direction favour the appearance of long pentagonal nanowires \cite{GarciaMochalesN08,GarciaMochalesJNM08,GarciaMochalesPSSA08}, and that there exists an optimal temperature at which the pentagonal nanowire yield is maximized \cite{GarciaMochalesPSSA08}.

In Refs. \cite{GarciaMochalesN08,GarciaMochalesJNM08,GarciaMochalesPSSA08} a method based on the time that the breaking nanowire lasts with a cross section $S_m\sim5$ (in units of atoms) was used to detect the formation of -5-1-5- structures. The quantity $S_m\sim5$ is close to the minimum cross-section of a pentagonal ring. As it is shown in Fig. \ref{fig1}b, the formation of the pentagonal nanowire is reflected in the curve of the nanowire minimum cross-section versus time, presenting a long plateau around $S_m\sim5$ during the pentagonal nanowire formation. A large statistical occurrence of pentagonal nanowires (for a given initial size, orientation and temperature) is reflected in its histogram of minimum cross-section $H(S_m)$ as a huge peak centred at $S_m\sim5$ (as it is shown in Fig. \ref{fig1}c. In those previous works, the relative height of that peak or its area has been used as a parameter to classify the conditions and probability of formation of icosahedral nanowires, as well as the characterization of individual pentagonal nanowires was made measuring the length of that cross-section plateau.

However, this measure only provides a qualitative value of the average length of the pentagonal nanowires formed under stretching, and provides little information about the actual lengths of such pentagonal chains. Deformations of non-pentagonal regions during the stretching or the existence of stages on the formation of the pentagonal nanowire that provide a cross-section value distant from $S_m=5$ lead to the miscalculation of the length of the pentagonal region. This method can not determine either the number of pentagonal rings that form the tubular structure. In order to overcome these limitations, in this paper we present an algorithm that allows the automatic identification of pentagonal rings structures as well as the determination of the actual pentagonal nanotube length $L_p$. With this new tool we have revisited the Ni case case complementing previous statistical analysis.

\section{Computational Methods}

In our study we have followed the same approach of previous papers on Ni to perform the simulated nanowire breakage. In short, we have used semi-classical Molecular Dynamics (MD) methods at constant temperature to study the structure and rupture of metallic nanowires. The atomic interactions are described by the Mishin {\it et al.} parameterization \cite{MishinPRB99} of the embedded atom method (EAM) potentials \cite{DawPRL83,FoilesPRB85}. A nanowire breakage simulation starts with a parallelepiped of atoms ordered according to a FCC structure with bulk Ni lattice parameter $a=3.52$\AA. The nanowire is stretched along the $z$ coordinate at constant velocity till the nanowire breaks. During the stretching process the accurate knowledge of the atomic coordinates and velocities allows the characterization of the geometry (in particular the minimum cross section $S_m$ \cite{BratkovskyPRB95,SorensenPRB98}), forces, and kinetic and potential energies of the stretched nanocontact. The procedure of the breakage simulation and the $S_m$ calculation in atoms units are described in detail in Refs. \cite{GarciaMochalesN08,GarciaMochalesJNM08,GarciaMochalesPSSA08,GarciaMochalesMSP05,GarciaMochalesAPA05,HasmyPRB05}.

We are interested in this study on the appearance and identification of icosahedral structures, so we have focussed on in nanowires stretched along the [100] direction due to their large probability to form pentagonal nanowires \cite{GarciaMochalesN08,GarciaMochalesJNM08,GarciaMochalesPSSA08}. The number of atoms of the initial nanowire is 1029 (21 layers $\times$ 49 atoms per layer) and the temperatures used were $T=4, 160, 225, 300, 375, 465, 550, 690$ and $865$K (bulk Ni melting temperature $T_m=1730$K).

In previous works pentagonal staggered nanowires were tracked by inspecting the minimum cross-section trace in the region $S_m(t)\sim5$; we define $\Delta t_5$ as the time spent by the nanowire in the $4 < S_m(t)< 6$ region (see Fig. \ref{fig1}b). This procedure presents two problems. First, it did not allow an exact identification of the simulations creating pentagonal structures. Second, as it is discussed in the introduction, that parameter, multiply by the stretching velocity, could not reflect the actual length of the icosahedral nanowire. To overcome these difficulties we have developed an algorithm that identifies the pentagonal structures that form the icosahedral nanowire. The algorithm also allows us to define its length $L_p$ as the distance between the outermost pentagonal rings and to count the number of pentagonal rings $n_p$ that form it.

The algorithm is based in the determination of the angular distribution of the nearest-neighbors atoms and provides a parameter ($\alpha(z)$) which compares the angular distribution of the projected nanowire atomic coordinates with that corresponding to a perfect pentagonal nanowire. For a given $z$ coordinate we consider a slice perpendicular to the $z$ (stretching) direction with a thickness of 2 \AA\ and centred on the $z$ value. The $N_t$ atoms inside such slice are projected onto the $xy$-plane, each one getting new 2D coordinates ${\vec \rho}_i$; then the centroid of this structure is calculated ${\vec \rho}_0 = \sum {\vec \rho}_i/N_t$.  The angular distribution is calculated from the angles $\varphi_{i,j}$ between the pairs of vectors ${\vec \rho}_i^{\ \prime}$ and ${\vec \rho}_j^{\ \prime}$ defining the projected atomic coordinates with respect to the centroid (${\vec \rho}_i^{\ \prime} = {\vec \rho}_i - {\vec \rho}_0$). The parameter $\alpha$ is calculated as
\begin{equation}
\alpha = \frac{2}{N_a}\sum_{i,j}^{N_a}\frac{\left | \varphi_{i,j}-m\varphi_0 \right |}{\varphi_0},
\end{equation}
where $N_a$ is the number of pair of atoms considered, $\varphi_0 = \pi/5$ is the reference angle of a perfect staggered pentagonal structure, and $m$ is the integer that minimizes the expression $| \varphi_{i,j}-m\varphi_0|$. To avoid spurious contributions from atoms near the centroid, only vectors satisfying  $|{\vec \rho}_j^{\ \prime}| > a/4$  are considered, i.e., center atoms are excluded from the calculation of $\alpha$. This algorithm is applied along the $z$-coordinate of the nanowire, displacing the imaginary slab $\delta z=0.1$\AA\ at a time. This results in a $\alpha(z)$ pentagonal profile of the nanowire. 

In order to minimize artifacts, the $\alpha(z)$ curve is softened over a wider window. This softened curve $\langle \alpha\rangle(z)$ is defined as 
\begin{equation}
\langle \alpha\rangle(z) = \frac{1}{\Delta_z}\int^{z+\Delta_z/2}_{z-\Delta_z/2}\alpha(z^{\ \prime})dz^{\ \prime},
\end{equation}
where a value of $\Delta z=1$\AA\ has been found appropriate. This average of $\alpha$($\langle \alpha\rangle$) over a 1\AA\ interval provides a quantity that distinguishes between pentagonal and non-pentagonal structures through the nanowire. We have observed that if the parameter $\langle \alpha\rangle(z)< 0.5$, the set of atoms around $z$ forms a structure similar to that of a pentagonal ring. On the contrary, if $\langle \alpha\rangle(z)>0.5$ the set of atoms presents another structure (bulk like -FCC-, helical or or disordered). 

\begin{figure}[tb]
\includegraphics[width=8.5 cm]{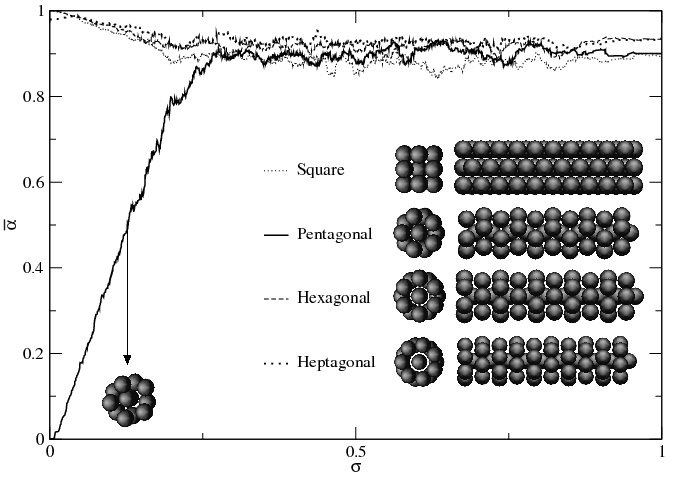}
\caption {\label{fig2} The average of the $\alpha$-parameter (${\bar\alpha}$) versus different strengths of the disorder parameter $\sigma$ (see text) for four test configurations: square 5-4-5-4, staggered pentagonal 1-5-1-5, staggered hexagonal 1-6-1-6 and staggered heptagonal 1-7-1-7 nanowires (the inset shows the perfect ordered configurations of the four test nanowires). $\sigma$ is the mean atomic displacement of atoms with respect to the perfect position of the ordered configuration. The average value of $\alpha$ for disordered nanowires was obtained averaging over 50 configurations.}
\end{figure}

To check the ability of the algorithm to discriminate between different structures in Fig. \ref{fig2} we show the average $\alpha$ values obtained using the algorithm along different nanowires with increasing amount of disorder. The test structures for the algorithm were: square nanowires with atoms sequence 5-4-5-4 taken from a FCC structure along the [100] direction; staggered pentagonal nanowires with atoms sequence 1-5-1-5; staggered hexagonal 1-6-1-6 nanowires; and staggered heptagonal nanowires with 1-7-1-7 sequence. The degree of disorder was measured with the mean displacement $\sigma$ of the atoms from their position in the perfect structure: $\sigma = \sum_i {\vec \rho}_i - {\vec \rho}_{i,0}/RN_t$) where  ${\vec \rho}_{i,0}$ corresponds to the ordered structure atom position, and $R$ is the effective radius of the ordered test configuration. The initial ordered structures ($\sigma =0$) are depicted as inset in Fig. \ref{fig2}, and only in the case of pentagonal nanowire the parameter $\alpha$ takes value 0, being $\sim$1 for the other structures. As the disorder increases, the $\alpha$ average (${\bar\alpha}$) varies: it increases for pentagonal nanowires and slightly decreases for the other nanowires. If the disorder with respect the initial structure is strong enough, the average of $\alpha$ for all the test nanowires converges to a value $\sim$0.9. 

\begin{figure}[tb]
\includegraphics[width=8.5 cm]{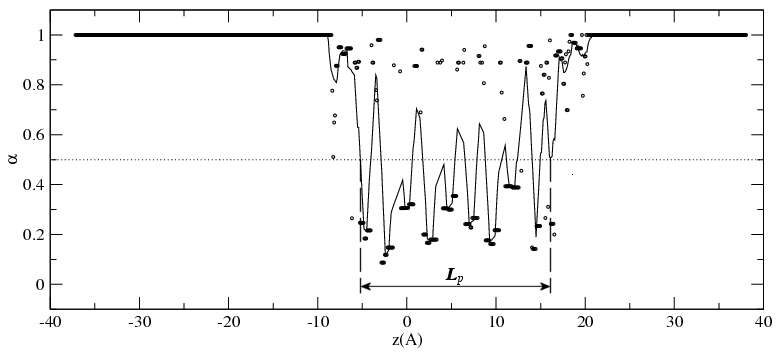}
\caption {\label{fig3} $\alpha$-parameter (dots) and its average $\langle \alpha\rangle$ (solid line) along the Ni[100] simulated nanowire at $T=375$K shown in Fig. \ref{fig1}a. For this particular case we include the icosahedral nanowire length $L_p$(=21.1 \AA). The doted line $\langle \alpha\rangle=5$  is the limit value we have chosen for the identification of pentagonal structures. The minimum values of $\langle \alpha\rangle$ correspond to the position of pentagonal rings forming the  icosahedral nanowire.}
\end{figure}

 In Fig. \ref{fig3} we present the result of the algorithm applied over a simulated nanowire that presents an icosahedral structure (that showed in Fig. \ref{fig1}a. Here we show both $\alpha(z)$ and the softened $\langle \alpha\rangle(z)$ curves. As illustrated in the figure, the algorithm returns value near to 1 when is applied to the ordered regions of the nanowire, and values below 1 for the thinnest part of the nanowire. Minima of $\langle \alpha\rangle$ correspond to the position of the pentagonal rings; as they are not perfect ordered structures (but still they can be recognized as pentagons) their $\langle \alpha\rangle$ values are greater than cero. We have chosen the value of $\langle \alpha\rangle=0.5$ as the limit value to recognized a pentagonal structure. As it can be seen in Fig. \ref{fig2}, non-pentagonal tubular structures (even with strong disorder) have a $\langle \alpha\rangle$ value higher than 0.5. The pentagonal nanowire, even with a relative strong disorder, presents a $\langle \alpha\rangle$ lower than 0.5; a disordered pentagon with $\langle \alpha\rangle>0.5$ can not be identified as a regular pentagon (see Fig. \ref{fig2}). The value $\langle \alpha\rangle=0.5$ discriminates between pentagonal and non-pentagonal structures. We define the pentagonal nanotube length $L_p(t)$, observed during stretching at a given time $t$, from the maximum and minimum $z$-coordinates with $\langle \alpha\rangle=0.5$ as it is shown in Fig. \ref{fig3}. $L_p^m$ is the maximum value of $L_p(t)$, observed when the nanowire is about to break, and $n_p$ is the number of pentagonal rings forming the icosahedral nanowire at its late stage (equivalent to the number of $\langle \alpha\rangle$ minima below 0.5).

\section{Results}

Once the ability of the algorithm to identify pentagonal structures has been checked, we apply the algorithm to our simulations on Ni [100] nanowires. As in previous works the aim is to carry out a statistical study for a broad range of temperature, obtaining probability distributions of pentagonal nanowires lengths $L_p^m$ and number of pentagonal rings $n_p$. This will give the optimal temperature required for maximizing their occurrence probability, taking into account that $n_p=2$ is the minimum value defining a pentagonal nanowire. We compare our results with those obtained using the time interval $\Delta t_5$ to characterize pentagonal nanowires.

\begin{figure*}[htp]
\centering
\includegraphics[scale = 0.2]{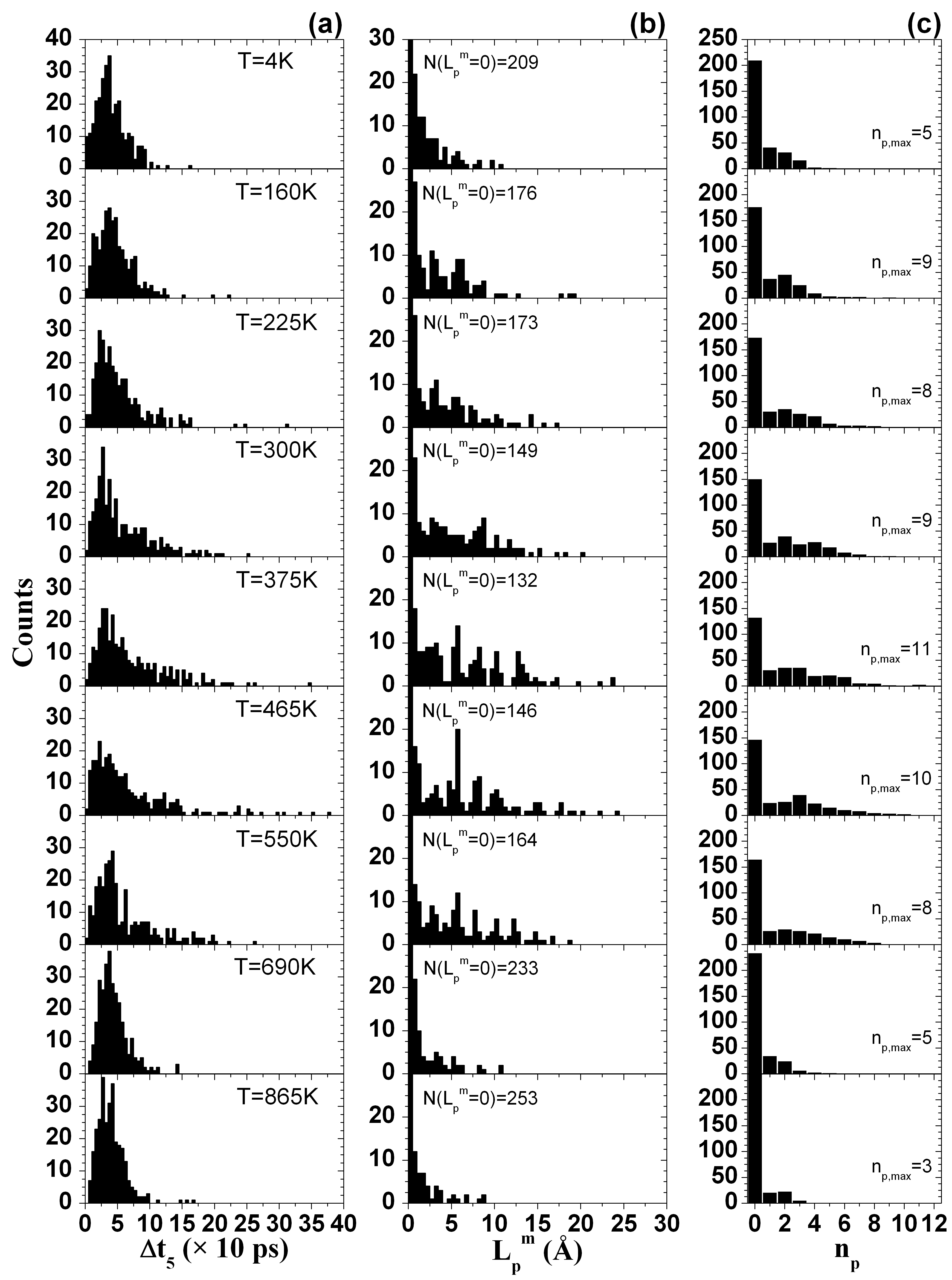}
\caption {\label{fig4} Distribution function for different magnitudes obtained from a set of 300 Ni[100] simulated nanowires with a size of 1029 atoms. (a) Time $\Delta t_5$ spend in the region of cross section $4.5 < S_m(t) < 5.5$ by the set of nanowires. (b) Maximum pentagonal nanowire length $L_p^m$; the value $N(L_p^m=0)$ corresponds to the number of nanowires from the set of 300 that do not show any pentagonal ring in its structure. (c) Number of pentagonal rings $n_p$ just before the pentagonal nanowire breakage (associates with the length $L_p^m$). Each row corresponds to a different temperature: 4, 160, 225, 300, 375, 465, 550, 690 and 865 K respectively.}
\end{figure*}

In Fig. \ref{fig4} we show the distribution (obtained from 300 simulations for each temperature) of $\Delta t_5$, $L_p^m$ and $n_p$ for a set of temperatures ranging from near 0 to $T_m/2$. 

The $\Delta t_5$ distributions show a smooth shape, being the most remarkable differences with the temperature the presence of a long tail for the cases ranging from 300 to 550 K. The initial part of the distribution (up to the maximum position around 30 ps) is due to the plastic deformation of all the nanowires, and it does not reflect necessarily the formation of any icosahedral structure. Higher values of $\Delta t_5$ would reflect the creation of long structures with $S_m\sim5$, (pentagonal nanowires) and the distribution tail would reflect the length of that structures and probability of findding one in a stretching proccess. In cases with a long tail distribution (300-550 K) the pentagonal structure in some nanowires seems to last up to 350 ps, that is equivalent to structures with a length of 14 \AA\ long. 

The way $\Delta t_5$ is obtained is similar to that used to measure experimental plateau lengths distributions \cite{RubioPRL01}. However, we have found that this quantity does not reflect the actual maximum length of the icosahedral nanowires $L_p^m$ . We will return to this issue later, but it is clear when $\Delta t_5$ and $L_p^m$ distributions are compared.

The length $L_p^m$ distributions (Fig. \ref{fig4}b) present a clearly peaked shape, with the peaks separated approximately by integer values of the distance $d_{5-5}=2.22$\AA, the calculated equilibrium separation between successive staggered pentagonal rings \cite{GonzalezPRL04,SutrakarJP08}. The distribution value for $L_p^m =0$ corresponds to those cases where the nanowire does not form any pentagonal structure (np=0). Those cases with $0<L_p^m <2$\AA\ correspond to nanowires showing  a unique pentagonal ring ($n_p=1$). We consider that the nanowire has formed a pentagonal or icosahedral nanowire when $n_p\ge2$, i.e. when it contains at least a full icosahedron in its narrowest part. In agreement with the $\Delta t_5$ distributions, the $L_p^m$ distributions from $T=300$K to 550K show large tails corresponding to those pentagonal nanowires including a large number of pentagonal rings. In particular, we have detected two cases with $n_p=11$ pentagonal rings for the ensemble of breaking events studied at $T=375$K set (both pentagonal nanowires were characterized with $L_p^m=22.2$\AA\ and 23.7\AA, respectively). 

In Fig. \ref{fig4}c is shown the np distribution for the same sets of simulations at different temperatures. Clearly, for all the temperatures, the most probable case is that the nanowire does not form any pentagonal structure, ie. $n_p=0$. ($N(n_p)=0$ agrees with that of $N(L_p^m =0)$). However for the intermediate range of studied temperatures it is most probable to form a two pentagonal rings structure than that only showing a single ring. Indeed, for two temperatures (375K and 465K) the most probable nanowire structure possesses 3 staggered pentagons ($n_p=3$). These two temperatures also exhibit the longest icosahedral structures. Ideally every structure with a $n_p\ge2$ would give a sharp peak in the $N(L_p^m)$  distribution. But real structures show tilted pentagonal rings, there is not perfect aligned with $z$-axis and the pentagonal rings separation is longer that the equilibrium, producing broad peaks in the $N(L_p^m)$ figure.

\begin{figure}[tb]
\includegraphics[width=8.5 cm]{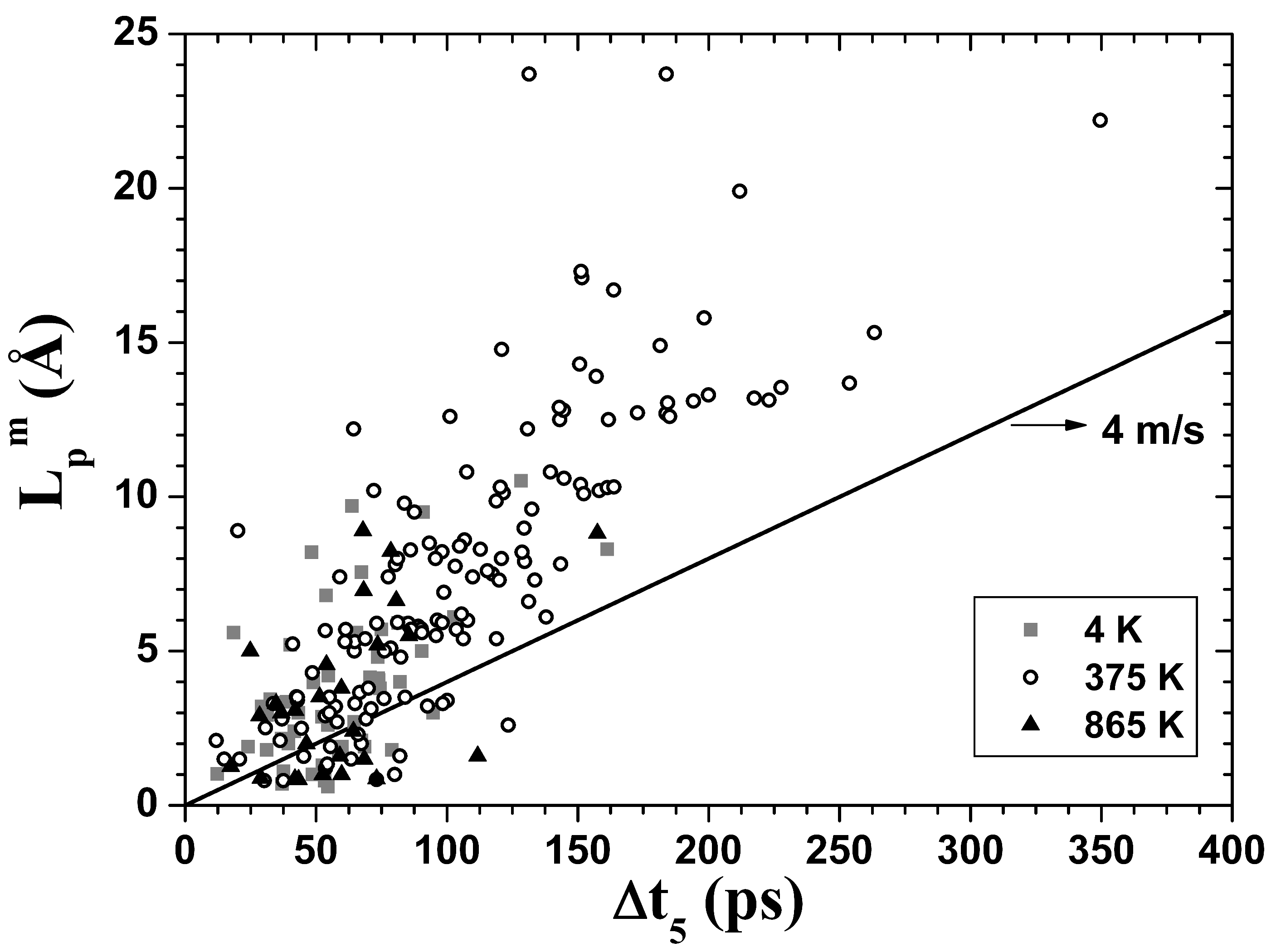}
\caption {\label{fig5} The $L_p^m$ length versus its corresponding $\Delta t_5$ time for the Ni[100] simulated nanowires that present the formation of an icosahedral nanowire, i.e. $n_p\ge2$ (at $T=4$, 375 and 865 K). The solid line is the expected relation between the nanowire length increasing and time stretching at a constant velocity of 4m/s. }
\end{figure}

As it has been pointed out previously, the parameter $\Delta t_5$ does not reflect the actual length of the pentagonal nanowire, and its distribution can not been taken as indication of the actual length distribution. ÕA prioriÕ, the shortest $\Delta t_5$ values must correspond to nonpentagonal nanowires. However, for larger $\Delta t_5$ values, their maximum length $L_p^m$ would be shorter than the nanowire length increase calculated as $\Delta t_5$ times the stretching velocity. This could be due to the plastic deformation suffer by different regions of the nanowire during the stretching process. In Fig. \ref{fig5} we show the $L_p^m$ value against the corresponding $\Delta t_5$ value for the cases where the stretched nanowire creates a pentagonal nanowire (i.e, the plotted data are restricted to nanowires with $n_p\ge2$). Fig. \ref{fig5} also depicts the expected increased length as function of the time (for clarity only results from three temperatures are shown; others temperatures shown a similar distribution. This figure demonstrates that the former argument is wrong; since the maximum icosahedral nanowire length is generally larger than that expected froma a direct measurement of $\Delta t_5$. 

Fig. \ref{fig5} proves that, in general, the $\Delta t_5$ measure underestimates the maximum length for long icosahedral nanowires. For the case of short nanowires $\Delta t_5$ tends to overestimate the actual pentagonal chain length. The magnitude of such underestimation depends on the particular nanowire. The cause of the underestimation is that, in many nanowires, when the minimum cross-section starts to get close to $S_m\sim5$, there is already formed an icosahedal structure with several pentagonal rings (generally two or three) (see Fig. \ref{fig2}b on reference \cite{GarciaMochalesPSSA08} for an example). The previous proto-icosahedral structure shows $S_m>6$ and, after stretching, it becomes a ÒclassicalÓ pentagonal nanowire completely formed showing the -1-5-1-5-structure with $S_m\sim5$. From this moment, and upon increasing stretching, the pentagonal nanowire adds new pentagonal rings to the structure. Depending on how the formation process occurs in every particular event, the $\Delta t_5$ measure lacks to take into account the length of several pentagonal rings, and therefore underestimated the total icosahedral nanowire length. So the pentagonal nanowire length estimations of previous studies \cite{GarciaMochalesN08,GarciaMochalesJNM08,GarciaMochalesPSSA08} reflected shorter lengths than the real ones. Only the direct measure of the length (applying the algorithm described in this paper or similar methods) can provide information about the formation of pentagonal chains, allowing the statistical study of the appearance of pentagonal nanowires.

\section{Conclusions}

We have present a computational method to identify and characterize icosahedral nanowires. This methodology also allows the determination of the pentagonal chain length as well as the number of pentagonal rings that forms it. We have tested the proposed algorithm for several ad-hoc ordered and disordered structures, proving that it can satisfactorily distinguish staggered pentagonal nanowires from other tubular structures.

The new methodoloy has been applied for statistically studying hundreds of Ni[100] breaking nanowire simulations at different temperatures (ranging from 4K to 865K), obtaining the pentagonal length $L_p^m$ and number of rings np distributions. We have compared these results with the $\Delta t_5$ distributions, already used in previous works to identify the existence of pentagonal nanowires. We have shown that the quantity $\Delta t_5$ generally underestimated the length of the icosahedral nanowire and, as consequence it is not adequate for the characterization of icosahedral nanowires.

The $L_p^m$ distributions show a peaked structure, similar to those found for linear atomic chains \cite{RubioPRL01} with clear peaks separated by integer values of the calculated equilibrium distance between two consecutive staggered pentagons. The $n_p$ distribution shows that the formation of icosahedral structures is not a favorable situation for low and high temperatures. However, there is a range of temperatures (300-550 K) with a large probability (above 50\%) of obtaining long pentagonal nanowires ($n_p\ge2$) from stretching processes. This temperature dependence of the formation of pentagonal chains opens a technological way to optimize the fabrication of these nanoobjects.

\section{Acknowledgments}

This work has been partially supported by the Spanish DGICyT (MEC) through Projects FIS2005-05127, BFM2002-01167-FISI, and FIS2006-11170-C02-01, and by the Madrid Regional Government through the Programmes S-0505/MAT/0202 (NanoObjetos-CM) and S-0505/TIC/0191 (Microseres-CM). One of us (PGM) also acknowledges Spanish MEC by the financial support through its ÒRam\'on y CajalÓ Programme.


\end{document}